\documentclass[11pt]{article}
\usepackage{amsmath, amssymb, amscd, amsthm, amsfonts}
\usepackage{graphicx}
\usepackage{hyperref}
\usepackage{cite}
\usepackage{authblk}
\usepackage{lmodern}
\usepackage{siunitx}

\usepackage[T1]{fontenc}
\usepackage[utf8]{inputenc}
\usepackage{textcomp}

\UseRawInputEncoding

\title{\fontsize{16}{23}\selectfont\bfseries Inverse-design of two-dimensional magnonic crystals\\ via topology optimization\\ with frequency-domain micromagnetics
}

\author[1]{Ryunosuke Nagaoka}
\author[1,2]{Takahiro Yamazaki}
\author[3]{Chiharu Mitsumata}
\author[4]{\vspace{1em}\\ Yuma Iwasaki}
\author[1]{Masato Kotsugi}

\date{}

\affil[1]{Department of Materials Science and Technology, Tokyo University of Science, Tokyo 125-8585, Japan} 
\affil[2]{Faculty of Engineering, Yokohama National University, Yokohama 240-8501, Japan} 
\affil[3]{Graduate School of Pure and Applied Sciences, University of Tsukuba, Tsukuba 305-8571, Japan} 
\affil[4]{Center for Basic Research on Materials (CBRM), National Institute for Materials Science (NIMS), Tsukuba 305-0047, Japan}

\begin{document}

\maketitle

\begin{abstract}
Magnonic crystals (MCs) are emerging spintronic metamaterials capable of manipulating transmission properties of magnons, the quanta of spin waves. Due to the complex relationship between lattice geometry and magnonic band dispersion, it remains challenging to establish general design strategies for optimizing targeted properties in MCs. In this study, we demonstrated an inverse-design framework for two-dimensional MCs to explore unconventional lattice structures with large magnonic band gaps. We employed genetic algorithms to enable global exploration of structures with a complete band gap as the objective property, and used frequency-domain micromagnetic simulations for computationally efficient band gap evaluation. Our established inverse-design method successfully discovered several previously unreported designs of MCs, whose performance was validated using time-domain micromagnetic simulations. Furthermore, we observed that the design landscape becomes increasingly non-convex at high-order bands, suggesting the existence of multiple design solutions. The overall inverse-design framework is expected to be broadly applicable to experimentally accessible material systems and device dimensions, facilitating the formulation of design rules for MCs.
\end{abstract}

\section{Introduction}

Spin waves (SWs) are collective magnetization excitations that propagate in a phase-coherent manner and have attracted considerable attention as information carriers for beyond-CMOS technologies \cite{Flebus2024-jg, Chumak2022-jv}. A wide range of applications has been proposed and demonstrated, including logic circuits \cite{Khitun2010-hv,Kanazawa2016-um} memory devices \cite{Fan2023-dn} physical neural networks \cite{Papp2021-wv}, and neuromorphic computing architectures \cite{Nagase2024-ob,Namiki2024-cz,Gartside2022-rm,Korber2023-xi}. Among these applications, magnonic crystals (MCs), in which magnetic hosting materials of SWs are periodically modulated in space, have been proposed as metamaterial structures for tailoring the dispersion relationship of magnons. Such periodic structures promote coherent SW scattering, which gives rise to designed magnonic band dispersions and wave functions. One- \cite{Chumak2009-tz}, two- \cite{Mori2024-qn}, and three-dimensional \cite{Krawczyk2008-yu} MCs have been investigated both experimentally and theoretically. In addition, the resulting characteristics are flexibly tunable through external magnetic fields and ferroelectricity. This tunability places MCs among the magnetic counterparts of photonic crystals \cite{https://doi.org/10.1002/anie.200907091}, phononic crystals \cite{Liao2024-vf}, and multiphysics metamaterials that couple different wave phenomena\cite{Burns2025-ok, Sun2025-zp}. \par
One approach for tailoring magnonic band dispersions in MCs is to control the geometry of patterned magnetic materials. For instance, geometrically engineered band dispersions, including nonreciprocity, flatband dispersions, and band gap formation has been achieved by introducing arbitrary structural geometries \cite{Prajapati2024-ji,Park2021-lf,De2021-dl,Yang2025-rz} and controlling lattice symmetries \cite{Klos2012-dc,Yang2011-lm}. Among the resulting functionalities, magnonic band gaps (MBGs), where propagation of SWs is prohibited along specific excitation frequency, have been one of the central design targets because they are highly tunable through geometry and straightforward to measure experimentally via microwave measurements \cite{Wang2010-ae}.\par
However, due to the intricate relationship between lattice geometry and magnonic dispersions, comprehensive design guidelines for enlarging MBGs and the optimal geometries for achieving the largest MBGs remain unclear. In most previous studies, the search for desired band dispersions has been confined to parametrically sweeping limited shape-related parameters, such as the scatterer radius and volume fraction, to identify optimal configurations \cite{Wang2013-ar}. Furthermore, those studies have primarily focused on lower-order bands (e.g., the first to third bands) for designing MCs with large MBGs opening, and higher-order bands, which may provide greater design flexibility, have therefore not been adequately explored. This is because higher-order bands are more sensitive to geometric modifications, which makes their behavior less predictable and complicates straightforward control, hindering the efficient design of MCs and the active application of higher-order magnonic modes.\par
In this context, one promising strategy for efficiently exploring optimal structures realizing desired properties is the inverse-design approach, in which target physical properties are defined as objective functions and parametrized design region are optimized through machine learning or mathematical optimization methods. This approach enables flexible exploration of structural configurations and significantly expands the design space for realizing nonempirical structural designs. Conventionally, the geometric inverse-design has been widely adopted in multiple types of metamaterials, including acoustic crystals \cite{Wu2002-em}, thermal metamaterials\cite{Jin2021-mu}, and optical metasurfaces \cite{Jin2019-rf}, but its application to magnonics has remained limited. \par
Recently, this inverse-design approach has begun to be incorporated into magnonic devices. For example, optimizing local magnetic field profiles has demonstrated experimentally and enabled improved control of demultiplexing characteristics and logic-operation accuracy \cite{Zenbaa2025-og,Zenbaa2025-mu}. In the calculation studies, optimizing spatial distributions of materials has enabled the design of SW lens \cite{Kiechle2022-en}, SW filter \cite{Yan2022-zt}, and frequency demultiplexers \cite{Wang2021-bx}. More recently, Voronov et al. demonstrated topology optimization using a level-set method and designed a high-efficiency frequency demultiplexer of SWs \cite{Voronov2025-lm}. This topology-optimization approach has been particularly effective in enabling flexible exploration of non-intuitive device designs. However, when evaluating magnonic band dispersions of 2D MCs as the objective function using time-domain micromagnetic simulations, each run has to be terminated at a finite cutoff time. As a result, obtaining broadband frequency responses with high spectral resolution requires long simulation times, which can be computationally expensive and may reduce the precision of band gap evaluation. \par

In this study, we demonstrate inverse-design topology optimization of 2D MCs by combining frequency-domain micromagnetic simulations based on the frequency-domain Landau-Lifshitz-Gilbert (FD-LLG) equation with a genetic algorithm (GA), a stochastic global optimization method. As the target property, we focus on complete magnonic band gaps (CMBGs) and perform nonempirical structural exploration using the GA. The use of FD-LLG simulations enables computationally efficient and reliable evaluation of magnonic band dispersions compared with time-domain simulations. In particular, we focus on design optimization in higher-order bands, which has not been extensively investigated. \par
Our objective is to identify previously unexplored MC designs exhibiting wide CMBGs. We further validate the robustness of optimized designs via broadband magnon excitation with time-domain micromagnetic simulation. Aiming to elucidate design rules behind the optimized structures, we conduct an unsupervised machine learning analysis to visualize design landscape of the optimized structures, extending an approach that we previously used to analyze Ising-like magnetic domain structures \cite{Nagaoka2024-jw, Masuzawa2026-nl}. The overall inverse-design approach is, in principle, applicable to more experiment-oriented conditions for efficient SW device design. We expect topology-optimized 2D MCs may open a pathway to greatly enhanced flexibility in magnonic band dispersion engineering beyond that available in conventional structural design.

\section{Methodology for inverse design of magnonic crystals}
In this study, we performed autonomous design exploration of two-dimensional magnonic crystals (2D MCs) by implementing an inverse-design loop based on iterative feedback calculations of objective functions (CMBGs). For evaluating CMBGs, we employ frequency-domain micromagnetic simulation framework, which is based on the FD-LLG equation and the finite element method (FEM)\cite{Zhang2023-et}. To begin with, we start from time-domain LLG equation. LLG equation is the equation of motion governing magnetization dynamics on nanometer length scales and picosecond time scales, containing a precession and a damping terms as follows:\\

\begin{equation}
\frac{\partial \mathbf{m}}{\partial t}
= -\gamma \mu_0 \, \mathbf{m} \times \left[ \mathbf{H}_{\mathrm{eff}}(\mathbf{m}) + \mathbf{h}(\mathbf{r}, t) \right]
+ \alpha \, \mathbf{m} \times \frac{\partial \mathbf{m}}{\partial t}
\label{eq:TD-LLG}
\end{equation}\\

where $\gamma$ is the gyromagnetic ratio, $\alpha$ is the Gilbert damping constant, $\mathbf{H}_{\mathrm{eff}}$ is the effective magnetic field. Magnonic band dispersions are typically estimated via time-domain micromagnetic simulations baed on LLG equation \eqref{eq:TD-LLG} combined with spatiotemporal fast Fourier transforms (FFT) \cite{Kumar2014-ri} or by semi-analytical approaches such as the plane wave method (PWM) \cite{Romero-Vivas2012-mt}. In contrast, FD-LLG simulations provide an efficient and reliable route to estimating dispersion relations for structures with arbitrary geometries represented by discretized finite elements. The FD-LLG equation is obtained by linearizing the magnetization dynamics assuming the small deviations from an equilibrium state, namely, the magnetization and effective field are decomposed into a static equilibrium background and a small dynamic perturbation as\\

\begin{equation}
\mathbf{m}(\mathbf{r},t)=\mathbf{m}_0(\mathbf{r})+\delta\mathbf{m}(\mathbf{r},t)
\label{eq:begin FD}
\end{equation}
\begin{equation}
\mathbf{H}_{\mathrm{eff}}(\mathbf{r},t)=\mathbf{h}^{\mathrm{eff}}_{0}(\mathbf{r})+\delta\mathbf{h}^{\mathrm{eff}}(\mathbf{r},t)
\end{equation}\\

where $\mathbf{m}_0$ and $\mathbf{h}_0^{\mathrm{eff}}$ are the equilibrium components, and $\delta\mathbf{m}$ and $\delta\mathbf{h}^{\mathrm{eff}}$ are the dynamic components of magnetization and effective field, respectively. There is an assumption that the excitation of moment is weak enough to consider linear approximation ($\delta\mathbf{m}\ll \mathbf{m}_0$ and $\mathbf{m}_0 \cdot\delta \mathbf{m}=0$). Under linear SW excitation, dynamic components of both the magnetization and magnetic field can be expressed in the form of harmonic functions using temporal Fourier transform as\\

\begin{equation}
\delta\mathbf{m}(\mathbf{r},t)=\delta\mathbf{m}(\mathbf{r},\omega)\,e^{i\omega t}
\end{equation}
\begin{equation}
\delta\mathbf{h}^{\mathrm{eff}}(\mathbf{r},t)=\delta\mathbf{h}^{\mathrm{eff}}(\mathbf{r},\omega)\,e^{i\omega t}
\label{eq:end FD}
\end{equation}\\

where $\omega$ is angular frequency of magnetization perturbation around its equilibrium states $\mathbf{m}_0$. This approximation enables the transformation of the time-dependent LLG equation into frequency-domain. By substituting Equation \eqref{eq:begin FD}-\eqref{eq:end FD} into Equation \eqref{eq:TD-LLG}, the FD-LLG equation is given as follows:\\

\begin{equation}
i\omega\delta\mathbf{m}
= -\gamma\mu_0(\mathbf{m_0} \times \delta\mathbf{h^{\mathrm{eff}}}
+ \delta\mathbf{m} \times \mathbf{h_0^\mathrm{eff}})
+ i\omega\alpha\mathbf{m_0} \times \delta \mathbf{m}
\label{eq:linearized LLG}
\end{equation}\\


By solving Equation \eqref{eq:linearized LLG}, it is possible to determine the spatial distribution of the magnetization perturbation for each frequency component. In this study, regarding the effective field terms, only the exchange field and external field were considered for static and dynamic component as $\mathbf{h}_0^{\mathrm{eff}}=A_{\mathrm{ex}}\nabla^2\mathbf{m}_0 + \mathbf{h}_0$ and $\delta\mathbf{h}^{\mathrm{eff}}=A_{\mathrm{ex}}\nabla^2\delta\mathbf{m} + \delta\mathbf{h}$. This approximation ensures that SWs are excited in the exchange mode. The MCs considered in this study have a lattice constant comparable to the exchange length, thereby the magnetostatic field can be neglected. 

\begin{figure}[htb!]
\begin{center}
\includegraphics[scale=0.8]{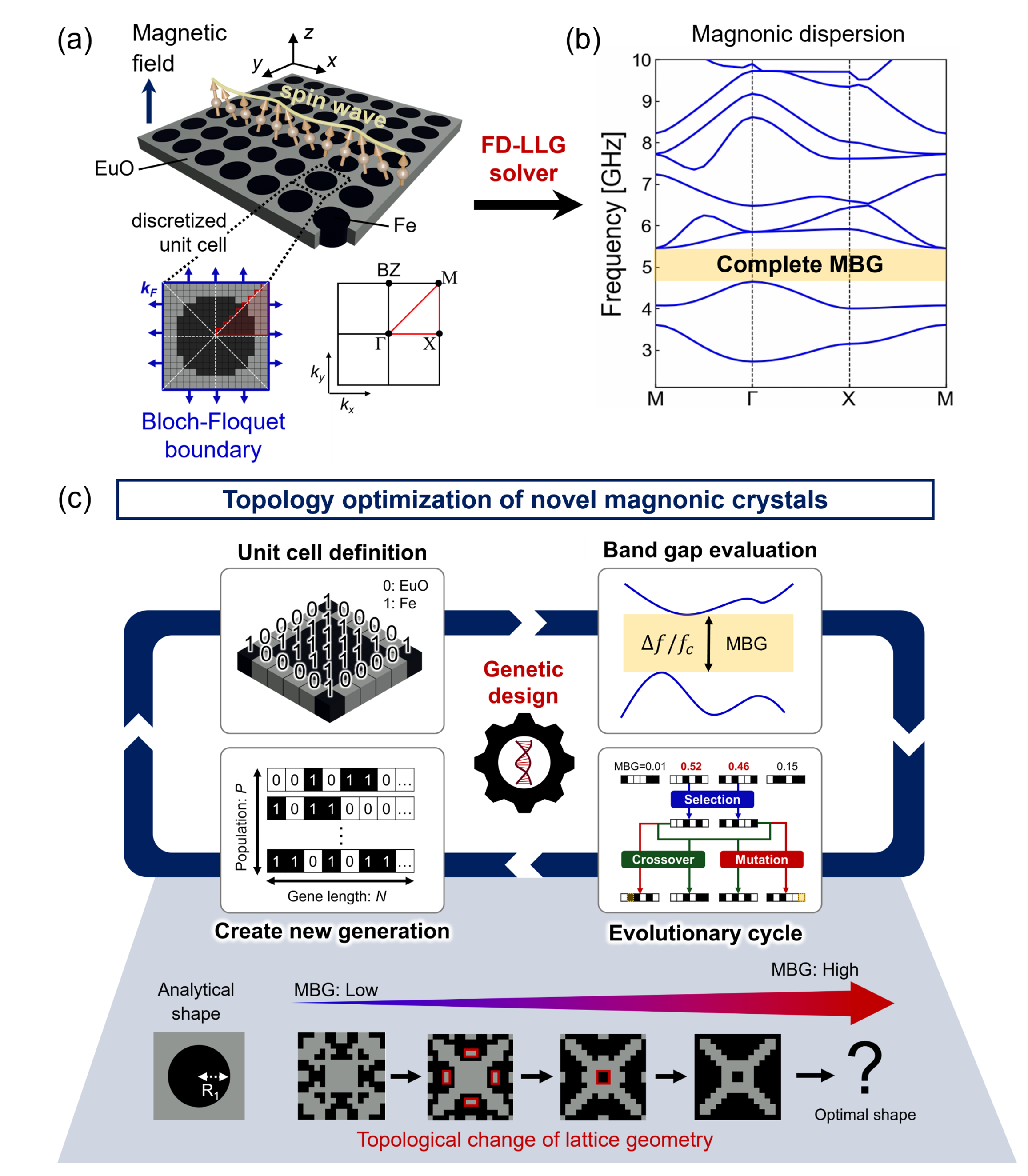}
\caption{a) Schematic illustration of square-circular 2D MC composed of EuO matrix and Fe circular dot. Spatially discretized unit cell structure of MC and the schematic image of 1st Brillouin zone are shown below. b) Corresponding dispersion relationship calculated on unit cell shown in a). c) Conceptual schematic of the inverse-design loop implemented in this study.}
\label{myfig1}
\end{center}
\end{figure}

Furthermore, by solving Equation \eqref{eq:linearized LLG} as an eigenvalue problem, it is possible to calculate the eigenmodes of $\delta\mathbf{m}(\mathbf{r}, \omega)$, enabling the calculation of the SW dispersion relation of MCs by scanning of wave vector $\mathbf{k}$. This was implemented by applying the Bloch-Floquet boundary condition in the periodic direction. The Bloch-Floquet condition allows periodicity in both the amplitude and phase of physical quantities, expressed by\\

\begin{equation}
\delta\mathbf{m}(\mathbf{r}+\mathbf{R})
=\delta\mathbf{m}(\mathbf{r})\cdot e^{-i\,\mathbf{k}_{\mathrm{F}}\cdot\mathbf{R}}
\end{equation}\\

where $\mathbf{k}_{\mathrm{F}}$ is a Floquet wave vector. In the calculation of magnonic dispersions in 2D MCs, this boundary condition are set in the $x$ and $y$ direction. This periodic approximation allows us to restrict the computational domain to a single unit cell. By solving the eigenmodes for each $\mathbf{k}_{\mathrm{F}}$, it is possible to acquire the SW dispersion in the form of $k-f$ diagrams. In this study, calculations were performed within the first Brillouin zone illustrated in Figure \ref{myfig1}a. While conventional numerical methods solve LLG equation in the time domain, a frequency-domain formulation offers significant advantages in both speed and accuracy for evaluating SW dispersions in complex systems including intricate geometries, which is advantageous for iterative optimization loop.


In the band gap evaluation, the discretized unit cell approximation was considered as shown in Figure \ref{myfig1}a. We selected a bicomponent MC composed of Europium oxide (EuO) and Iron (Fe) as the model system. This material combination is favorable for opening wide CMBG, our target property in this study, due to its high contrast in material parameters. We used a cuboid mesh with a side length of 2 nm, and the lattice constant of the MC was set to $a=32$ nm. In addition, as a boundary condition, Bloch-Floquet boundaries were applied in the $x$ and $y$ direction as shown in the Figure \ref{myfig1}a. Static magnetic field $\mathbf{h}_0$ was applied perpendicular to the film (in the $z$ direction), which was set to $\mu_0 h_0^z=$0.1 T. The static equilibrium magnetization $\mathbf{m}_0$ was assumed to be uniform in $z$ direction, thereby unit vector of magnetization was set to $\mathbf{m}_0=(0,0,1)$. The simulation parameters for individual materials were set as follows: for EuO, $A_{\mathrm{ex}}^{\mathrm{EuO}}=0.1\times10^{-11}$  J/m and $M_{\mathrm{s}}^{\mathrm{EuO}}=1.91\times10^6$ A/m; for Fe, $A_{\mathrm{ex}}^{\mathrm{Fe}}=2.11\times10^{-11}$ J/m and $M_{\mathrm{s}}^{\mathrm{Fe}}=1.752\times10^6$ A/m \cite{Cao2010-qf}. Since the analysis focuses on exchange mode SWs, only the exchange stiffness constant $A_{\mathrm{ex}}$ and the saturation magnetization $M_{\mathrm{s}}$ are considered as material-dependent parameters, because the analytical formulation of dispersion relation of exchange mode magnons in a single medium is given by $f(k)=\frac{\gamma\mu_0}{2\pi} [h_0 + \frac{2A_{\mathrm{ex}}}{\mu_0M_{\mathrm{s}}}k^2]$ \cite{Rezende2020-ex}. Damping constant was set to $\alpha=0.001$ for both materials. Furthermore, when defining material distribution, we defined only one-eighth of total area, utilizing the mirror symmetry, as indicated by red frame in the unit cell shown in Figure \ref{myfig1}a. This allows the definition of material distribution only in the region of interest, corresponding to 36 dimensions. Although this approximation limits the symmetry within square symmetry, it is reported that in exchange mode SWs, the largest gap can potentially be achieved when the scatterer has the same symmetry as that of lattice point\cite{Cao2010-qf}. Therefore, we focus on square symmetry; however, honeycomb lattices with sixfold symmetry are expected to yield a larger MBG, and should therefore be investigated in future work. The calculation described above was implemented by utilizing a third-party micromagnetic module provided in finite element simulation software COMSOL Multiphysics with the eigenvalue problem solved by solver package ARPACK \cite{Zhang2023-et}. \par

Figure \ref{myfig1}b shows the band dispersion calculated for the square-circular MCs illustrated in Figure \ref{myfig1}a using FD-LLG methods. The band structure reveals the emergence of a frequency range in which magnons are not excited in any propagation direction. This frequency range corresponds to a CMBG. We focus on this property as a model target to investigate the optimal geometry that maximizes CMBG in Fe-EuO system.\par
Figure \ref{myfig1}c illustrates the conceptual schematic of the design-optimization loop targeting to maximize CMBG of MCs. In this study, we adopted a binary encoding scheme and employed a GA as the optimizer. The GA is a well-established metaheuristic algorithm inspired by natural selection and biological evolution, which was originally developed in the field of bio-inspired computing \cite{Heiles2013-kq}. Because GA is a stochastic, gradient-free method, it is well suited to global optimization problems and is expected to discover unconventional MC geometries that realize large MBGs, which may be difficult to obtain by merely sweeping simple shape-related quantities. In the proposed inverse-design loop, the material distribution in a unit cell is represented by a binary vector (0/1) with the length of 36, which is regarded as a genetic sequence. The optimization proceeds iteratively by repeating the following steps: (i) definition of the unit-cell material distribution, (ii) evaluation of the band gaps using FD-LLG simulations, and (iii) generation of the next population by GA operations (selection, crossover, and mutation). (iv) Create bit arrays in new generation and redefine unit cell. Through this loop, the discretized cell structure is gradually transformed toward designs with larger band gaps. As an illustrative example, Figure \ref{myfig1}c shows a sequence of structural evolution during optimization. By iterating the loop, the initially discretized topology is progressively updated and converges to a non-intuitive topology that differs substantially from conventional analytical shape. Notably, during the evolution process, the emergence and disappearance of holes and isolated islands can be observed, indicating discontinuous topological changes. Such non-continuous topology transformations enable the proposal of previously unexplored MC topologies.\par
Next, we describe the algorithmic formulation of inverse-design process in detail. The optimization problem to maximize CMBG width is formulated as follows\\

\begingroup
\setlength{\belowdisplayskip}{50pt}       
\setlength{\belowdisplayshortskip}{0pt}  
\begin{equation}
\left\{
\begin{aligned}
&\text{maximize } C_n(\boldsymbol{\rho})
\qquad \text{w.r.t. } \boldsymbol{\rho} \in \{0,1\}^{N}
\quad (N = 36)
\\[6pt]
&C_n(\boldsymbol{\rho})
= \frac{\Delta f}{f_{\mathrm{c}}}
= 
\frac{
    \displaystyle \min_{k}:\, f_{n+1}(\mathbf{k})
    \;-\;
    \displaystyle \max_{k}:\, f_{n}(\mathbf{k})
}{
\frac{1}{2}
    [\displaystyle \min_{k}:\, f_{n+1}(\mathbf{k})
    \;+\;
    \displaystyle \max_{k}:\, f_{n}(\mathbf{k})]
}
\end{aligned}
\right.
\label{eq:objective func}
\end{equation}
\endgroup\\

    Here, $C_n(\mathbf{\rho})$ is a relative width of CMBG between $n$ th and $(n+1)$ th bands, an objective function defined in this study. $\mathbf{k}$ is wave vector, and $\Delta f$, $f_{\mathrm{c}}$ are absolute value of MBG and center frequency between band gaps, respectively. This relative band gap width is typically used for evaluating band gap width as a scale-invariant parameter in inverse design of photonic and phononic crystals \cite{Sigmund2008-dd,Han2019-rj}. $\mathbf{\rho}$ is a binary density variable of material with the dimension of $N$=36, which is allocated in each finite element cells. By using the elements of density variable $\mathbf{\rho}$, the material parameters are defined as $A_{\mathrm{ex}} (\rho_i)=(1-\rho_i ) A_{\mathrm{ex}}^{\mathrm{EuO}}+\rho_i A_{\mathrm{ex}}^{\mathrm{Fe}}$ and $M_{\mathrm{s}} (\rho_i)=(1-\rho_i ) M_{\mathrm{s}}^\mathrm{EuO} +\rho_i M_{\mathrm{s}}^{\mathrm{Fe}}$. When $\rho_i=0$, the $i$th cell is assigned to the parameters of EuO, whereas when $\rho_i=1$, it assigned to those of Fe. This means that there are $2^{36} \approx 10^{10}$ possible material distributions, making it impractical to find the optimal solution using exhaustive search. \par

\begin{figure}[htb!]
\begin{center}
\includegraphics[scale=1.1]{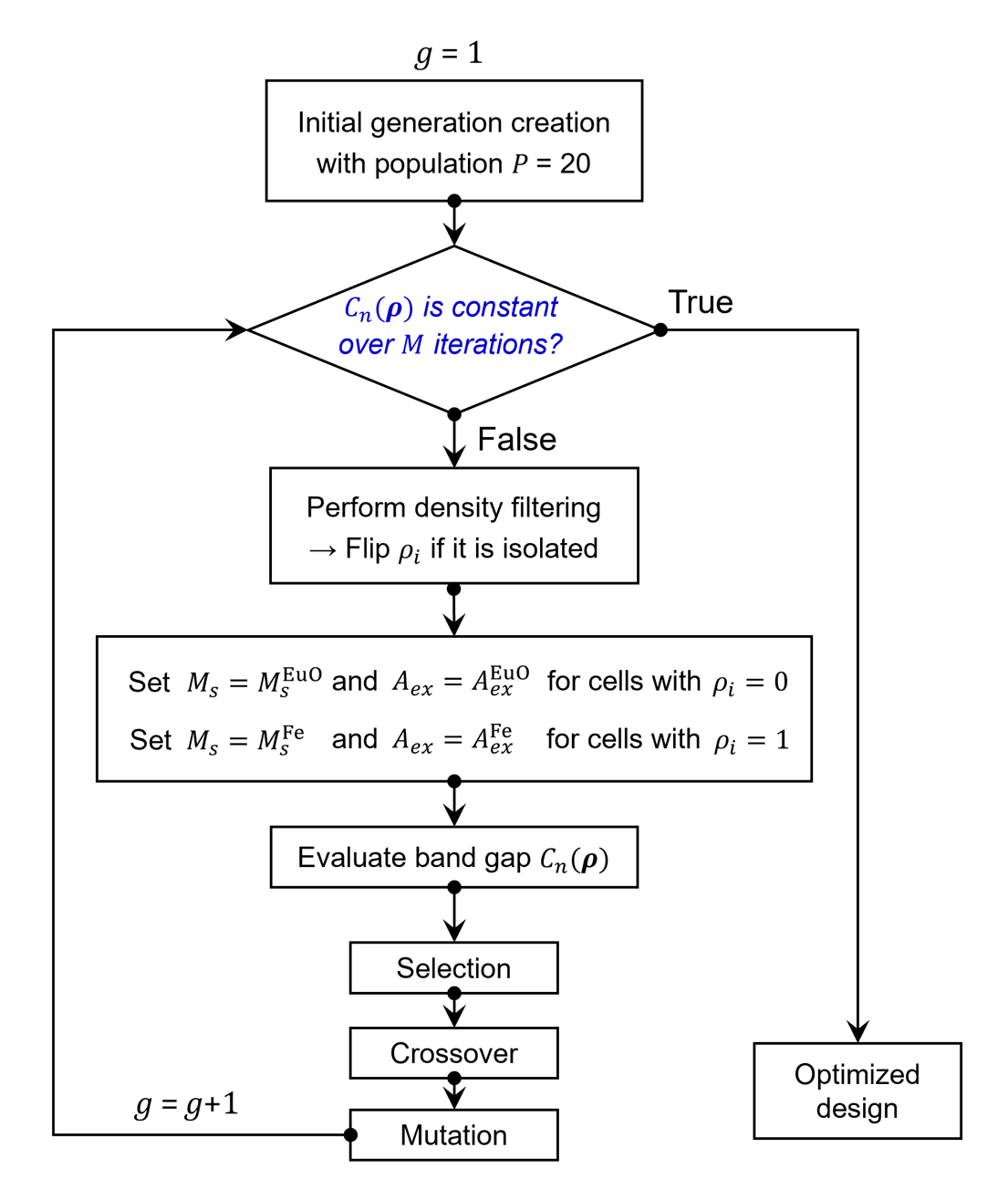}
\caption{Algorithm chart of design optimization process}
\label{myfig2}
\end{center}
\end{figure}
    
Figure \ref{myfig2} shows the algorithm chart of design optimization process. Firstly, the initial distribution was generated randomly. The number of generated feature vectors, which is called population, was set to $P$=20. With the binary-encoded 0/1 vector generated, it is possible to map material distribution onto each finite element cell. Then, the FD-LLG simulation was performed to evaluate the CMBG width for each design. Before band gap evaluation, we applied a density-filtering step to suppress spurious, noise-like features in the topology. Specifically, isolated bits whose nearest neighborhood predominantly consists of the opposite bits were flipped to match the surrounding bits. This filtering removes dot-like artifacts that are unfavorable for fabrication and improves the robustness of optimized designs. After evaluating relative width of CMBG, the natural selection process, a core process in GA, was performed to find candidate structure. The detailed process of GA is described in the Methods section.
After the sequence of GA, a new generation with the same number of $P$ is produced, and the same procedure is repeated until a convergence criterion is met; specifically, the optimization is terminated when the fitness value fails to improve for more than a prescribed number of generations $M$ (Here, we set $M=50$). Through this process, a candidate structure exhibiting a large CMBG is ultimately obtained. The whole optimization process explained above was implemented using MATLAB \cite{UnknownUnknown-lr}.\\

\section{Results and Discussions}
\subsection{Inverse-design demonstration}

\begin{figure}[htb!]
\begin{center}
\includegraphics[scale=0.9]{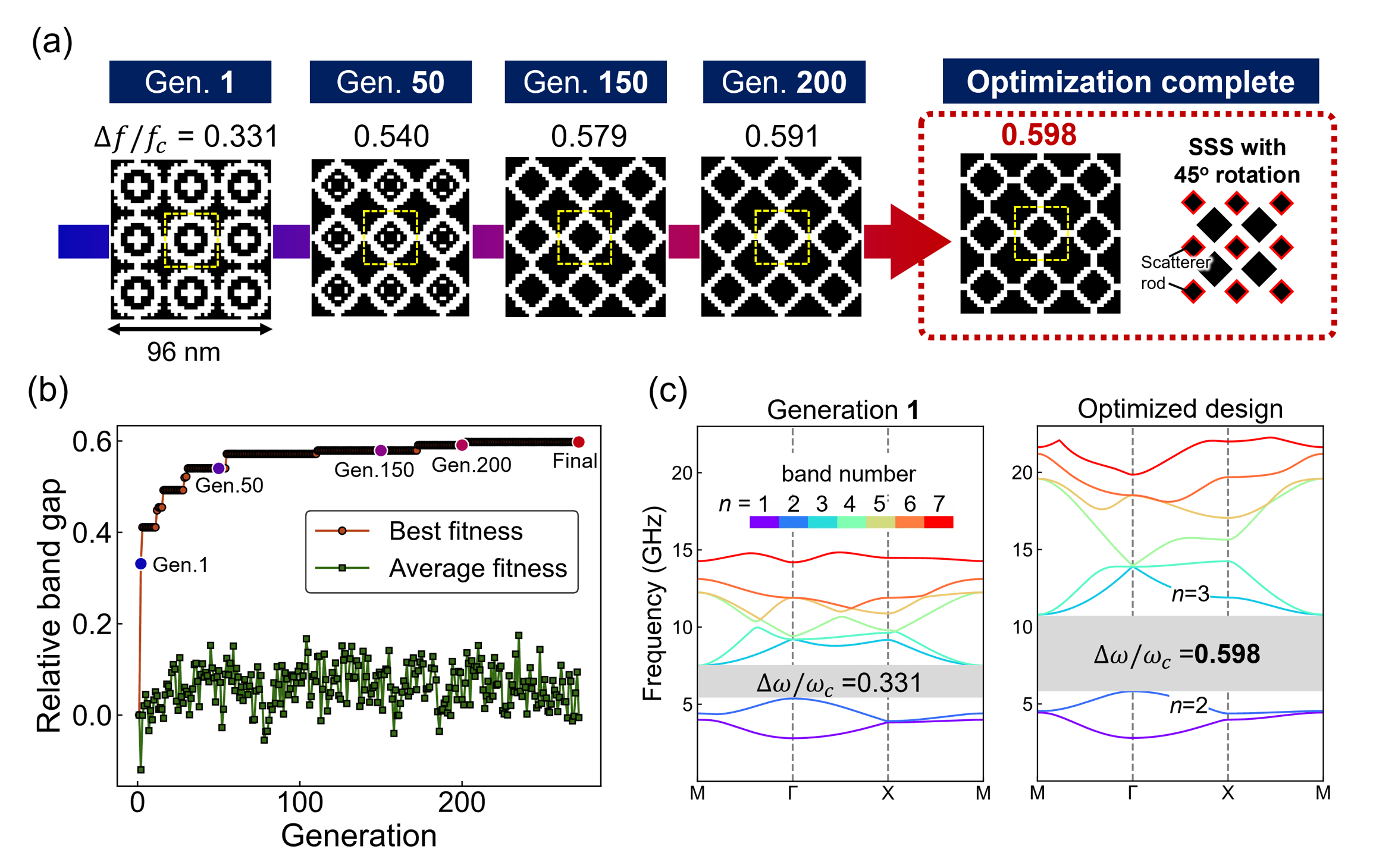}
\caption{a) Topological deformation of MC design during the optimization process maximizing CMBG between modes of $n=2$ and $n=3$. Note that the unit cell is expanded to $3\times3$ matrix for visualization. b) The history of change of fitness value (relative width of CMBG) along generation. c) Dispersion relationship of the best individual in 1st generation and final generation.}
\label{myfig3}
\end{center}
\end{figure}

As an initial demonstration of inverse-design, we explore the structures maximizing the relative CMBG between the modes of $n=2$ and $n=3$, which has been most extensively analyzed in previous studies. Figure \ref{myfig3}a shows the structural evolution of the MC during the optimization process. As the iterations proceed, the relative value of CMBG increases through a sequence of transformations, from $\Delta f/f_{\mathrm{c}}=0.331$ to $\Delta f/f_{\mathrm{c}}=0.598$ in the final generation. We see that the transformation process includes changes in connectivity and the generation and annihilation of holes or island structures, hence topological transformation was observed. Furthermore, the final optimized structure closely resembles the previously proposed square-square-square (SSS) structure with $45^\circ$ rotation of square rods. This geometry represents a promising candidate for opening a gap between the $n=2$ and $n=3$ modes, as the introduction of additional scatterers rods at face-centered positions in a square lattice is known to promote CMBG formation \cite{Wang2014-dh}. The $45^\circ$ rotation increases the filling fraction of scatterers within the unit cell, which is likely to contribute to the formation of a wide CMBG. Therefore, the optimal design was rediscovered by data-driven structural exploration, suggesting the effectiveness of global exploration using GA.\par
Another perspective to confirm the success of global optimization with GA is to see the trends in fitness values. As seen in Figure \ref{myfig3}b, while the best fitness in each generation is progressively improved through optimization, the average fitness of the population does not converge and continues to fluctuate. This indicates that, although the algorithm continues to search in the neighborhood of the current best solution, the diversity of population is preserved, allowing simultaneous exploration of other regions of the search space. Therefore, this behavior of average fitness suggests that global exploration was successfully achieved. \par


As a result, the data-driven structural exploration targeting the relative width of CMBG between $n=2$ and $n=3$ bands yields structures that are consistent with established theoretical understanding of wide-MBG MCs. This agreement supports the effectiveness of the present framework in identifying structures that has large CMBG between the $n=2$ and $n=3$ bands. However, the band dispersions calculated for the best individuals in the initial and final generations (Figure \ref{myfig3}c) further reveal that while the dispersions of the $n=1$ and $n=2$ bands remain largely unchanged, those of the high-order bands ($n\ge3$) are found to be sensitive to the geometric modulation. This means that the high-order dispersion relation has higher controllability by geometric designing but also indicates that there is a complex and nontrivial relationship between the lattice geometry and its resulting band structure. Such complexity has hindered the designing of MCs with large MBG at high-order bands. Hence, the design rules for MCs at high-order bands has not yet been systematically explored in prior studies.

\subsection{Exploring optimized designs in high-order bands}

\begin{figure}[htb!]
\begin{center}
\includegraphics[scale=0.8]{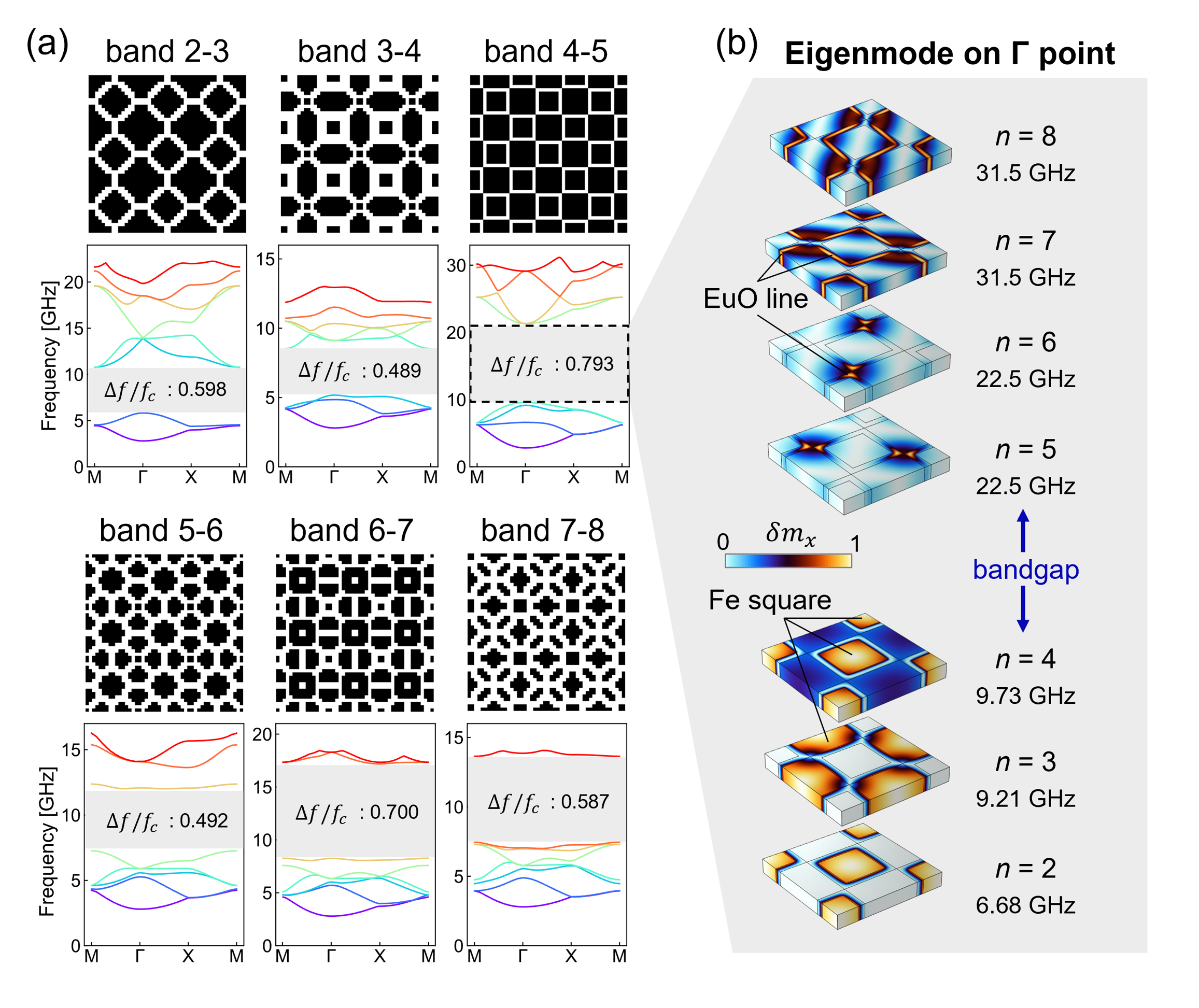}
\caption{a) Optimization results for different band numbers and corresponding dispersion relationships. b) The spatial mode profile of $\delta m_x$ at the $\Gamma$ point for the band 4-5 optimized structure.}
\label{myfig4}
\end{center}
\end{figure}

We next targeted the relative width of CMBG as an objective function and performed topology optimization at higher-order ($n\geq3$) bands. Hereafter, we refer to the structure optimized for the band gap between the $n$th and $(n+1)$th bands as the band-$n$-$(n+1)$ structure.
Figure\ref{myfig4}a shows the representative optimized structures for each band number and corresponding band dispersions. For each structure, it shows nontrivial MC designs that have not been previously proposed. Focusing on the structural trends along with band numbers, as the target band number increases, the structural features become progressively complicated and slim connections between the structural elements appear. The same trend appeared in the topology optimization of photonic crystals in TE modes\cite{Sigmund2008-dd}. This trend physically indicates the higher-order eigenmodes correspond to magnons with shorter wavelength, which are expected to resonate with structures of smaller characteristic length scales. \par


Focusing on individual designs, we find that the band-3-4 structure typically exhibit EuO lines crossing perpendicularly with oriented at $45^\circ$ with respect to the Floquet wave vector. Particularly intriguing is the band-4-5 structure, which consists of a simple arrangement of square Fe dots. To our knowledge, such a geometry has not previously been proposed as a large-MBG MCs, indicating that the present optimization reveals a previously overlooked structure in Fe-EuO MCs systems. For the band-5-6, band-6-7, and band-7-8 designs, the converged structures exhibit distinct topologies across different optimization runs. This indicates the difficulty in determining globally optimal geometries in high-order bands (a detailed discussion is given in Section \ref{sec:design landscape}). \par
In terms of mode analysis, we focus on the optimized band 4-5 structure, which has geometrically simple configuration and large CMBG width, and visualize the spatial distribution of the $x$ component magnetization perturbation $\delta m_x(\mathbf{r}, \omega)$. Since a band gap is formed at the $\Gamma$ point between band $n=4$ and $n=5$, the eigenmodes at the $\Gamma$ point are presented. The results show that, for bands below the lower edge of the gap, the eigenmodes are localized within the square Fe dots, whereas for bands above the upper edge of the gap, the modes are localized within the EuO line regions. This contrast in mode localization is likely responsible for the formation of the large CMBG in the optimized structure. \par
Compared with conventionally investigated structures, the optimized designs exhibit a greatly improved CMBG width. Specifically, we conducted an exhaustive comparison with various square-lattice MCs (shown in Supplementary Figure S2) and compared the relative band-gap widths using FD-LLG analysis. For the band-4-5, band-5-6, band-6-7, and band-7-8 designs, the relative CMBG widths are increased by approximately 90\%, 440\%, 170\%, and 830\%, respectively, compared with the best-performing conventional structures. Notably, for the band-3-4 design, CMBG between $n$=3 and $n$=4 is observed only in the topology-optimized structure, whereas none of the previously proposed structures exhibit a gap in this band range. These findings demonstrate the effectiveness of inverse-design structural exploration in high-order bands.\par

\par

\subsection{Evaluation of optimized designs with broadband magnon excitation }
\begin{figure}[htb!]
\begin{center}
\includegraphics[scale=0.95]{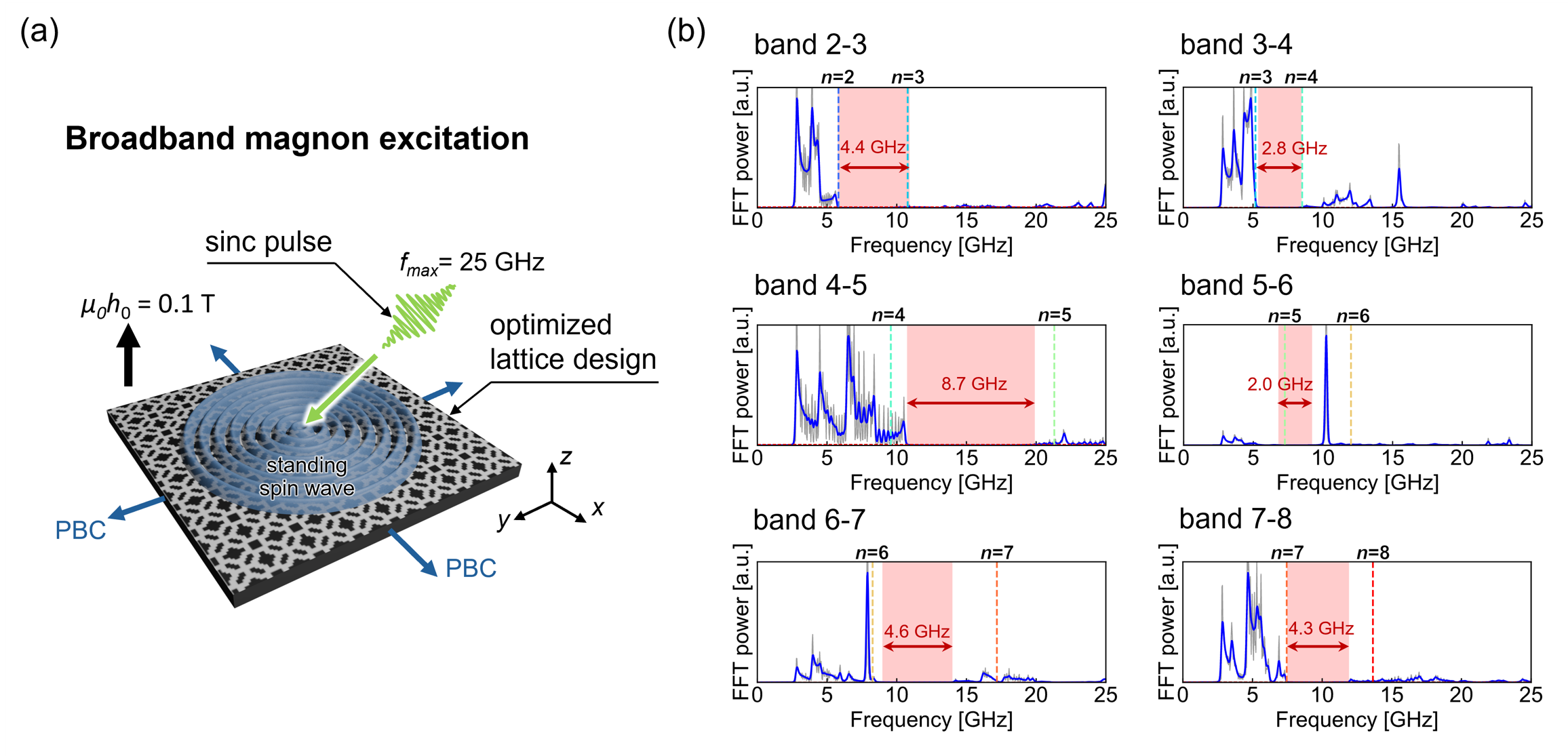}
\caption{a) Schematic illustrate of broadband magnon excitation using MuMax3 simulator. The sinc pulse of magnetic field was added in the extended super cell of optimized lattice. b) FFT spectrum of excited SW for optimized structures. The vertical dashed lines represent the upper and lower frequency bounds of the band gap obtained from FD-LLG analysis.}
\label{myfig6}
\end{center}
\end{figure}

To evaluate the propagation dynamics of SWs in optimized structures in finite dimensions, we created super cell from optimized unit cell and calculated excitation spectra of magnons using MuMax3 micromagnetic simulator \cite{Vansteenkiste2014-zv}. For spectral calculations, broadband magnon excitation is achieved by injecting a sinc pulse of magnetic field into the optimized lattice, as illustrated in Figure \ref{myfig6}a. The injected pulse is described by\\


\begin{equation}
H_{\mathrm{pulse}}(t) = \mu_0 H_{\mathrm{a}} \frac{\sin{[2\pi f_{\mathrm{max}}(t-t_0)}]}{2\pi f_{\mathrm{max}}(t-t_0)}
\label{eq:sinc pulse}
\end{equation}\\

The excitation waveform parameters were set to $\mu_0H_{\mathrm{a}}=0.01$ T and $f_{\mathrm{max}}=25$ GHz. The reference time was set to be $t_0=T/2$, where $T$ is the total excitation time of 30 ns. The unit cell was extended to 1.024 $\si{\micro\meter}$ $\times$ 1.024 $\si{\micro\meter}$ in the $x$ and $y$ directions to create super cell, and the periodic boundary condition (PBC) was set on the boundaries. The same cell size and material parameters as those used in the optimization loop are employed. The excitation spectra was obtained by performing a temporal FFT of the excited SW signals. \par

The calculated excitation spectra are shown in Figure \ref{myfig6}b. For all optimized structures, the presence of CMBGs is clearly confirmed by the existence of forbidden frequency gaps. Meanwhile, some bandpass frequencies with strong peak of FFT intensity can be observed, approximately corresponding to the band numbers. Notably, as expected from band dispersions, the band-4-5 design shows large absolute gap, which is measured $\Delta f=$8.7 GHz in width. It is worth noting that incorporating higher-order band gaps can yield larger band gaps that cannot be realized using only lower-order bands. For other geometries over $n=5$, they show $\Delta f=4.4-5.5$ GHz frequency gaps, showing superior band gap performance as well. In addition, we confirmed that even if the applied magnetic field s changed, the order trends of band number in absolute gap width do not change due to the linear shift of band dispersions. This might support the possibility of tuning frequency range while maintaining large CMBG in the optimized lattices.  \par
Overall, the results of time-domain magnon excitation support the existence of CMBGs and the validity of optimized MCs for manipulating SWs propagating properties. However, for high-order bands ($n\ge4$), the frequency of higher edge of band gap estimated in FD-LLG analysis (vertical dash lines in Figure \ref{myfig6}b) do not coincide exactly with frequency gaps of spectra. The high-mode bands are tend to be sensitive to boundary conditions, which might lead to unmatch with experimental results. While it remains necessary to consider boundary reflection and scattering of SWs and their effect in propagation properties, the frequency-domain inverse-design frameworks provides promising results offers a powerful tool for screening unconventional MC designs.

\subsection{Visualization of design landscape via dimensional reduction}
\label{sec:design landscape}

To evaluate the optimization process, we systematically performed inverse-design loops 12 times for each band number, using different random seeds to generate the distinct initial structures for each run, thereby reducing the dependence of the optimization results on the initial conditions. Because GA is a stochastic optimization algorithm and generates multiple design candidates over successive generations, we expect it to discover multiple design solutions, and to enable the visualization of global design landscapes. For visualizing the design exploration space, we performed unsupervised dimensionality reduction on all binary design arrays generated during optimization and mapped them into a 2D space. \par
\begin{figure}[htb!]
\begin{center}
\includegraphics[scale=0.9]{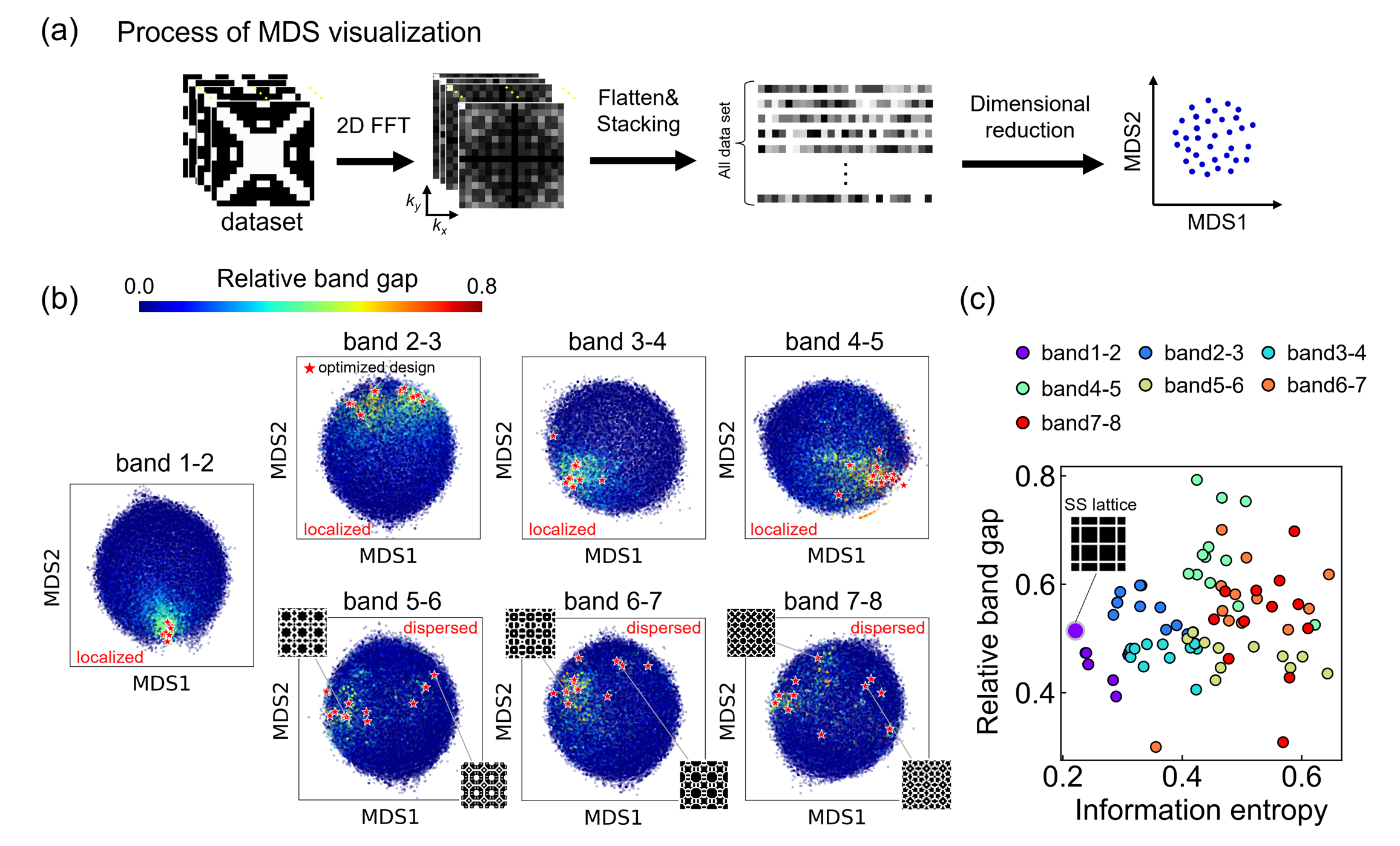}
\caption{a) MDS mapping of unit cell structures for each band numbers. Colormap represents relative width of CMBG, and the star marker represents the final optimal design. b) Correlation of information entropy calculated from unit cell structure and relative band gap. }
\label{myfig7}
\end{center}
\end{figure}

We employed combination of 2D FFT with multidimensional scaling (MDS), one of the dimensionality reduction techniques for 2D mapping. We expect 2D FFT is suitable for the current dataset, because the unit cell structure has translational symmetry in two dimensions, allowing the structural periodicity to be characterized via frequency-space features. MDS can represent high-dimensional data in a low-dimensional data space by preserving distances in original space, allowing high interpretability in clustering and similarity analysis. The detailed process is depicted in Figure \ref{myfig7}a. First, each 36-dimensional design vector was expanded into a unit-cell array ($16\times16$), and a 2D FFT was applied to obtain the Fourier spectrum and extract information of structural periodicity. The Fourier spectrum was then flattened into a 256-dimensional vector, and MDS based on Euclidean distances was performed to reduce the dimensionality from 256 to 2. For MDS calculation, scikit-learn, a Python package for machine learning, was utilized \cite{Pedregosa2011-qe}.\par
Figure \ref{myfig7}b shows the MDS mappings for each band number. The color map represents the relative band gap value of each structure, and each point corresponds to a unit cell geometry. For the band-1-2, band-2-3, band-3-4, and band-4-5 designs, finally optimized structures (marked by stars) form localized clusters in the reduced space. This indicates that the optimization process typically converges toward similar structures for lower bands. In contrast, for the band-5-6, band-6-7,and band-7-8 designs, the optimized structures are more broadly distributed, and the structural similarity between solutions is significantly lower. These results suggest that, in high-order bands, the optimization yields different solutions across runs and admits multiple local optima, implying that the design landscape is increasingly non-convex. The variation in optimized topology can be seen in the Supplemental Figure S1. From a physical perspective, these behavior can be understood that the high-order resonances can support multiple resonance modes because the increased degrees of freedom in finer geometries allow a larger number of possible configurations. \par

In addition, as a metric to quantify the complexity of the optimized geometries, we introduce the Shannon information entropy. Shannon entropy originally describes configurational entropy in Heisenberg spin models and utilized as a measure of geometrical complexity in metasurface structures \cite{Cui2016-mv}. 
Figure \ref{myfig7}c shows the correlation between the Shannon entropy of the final optimized structures and the corresponding band gap widths for each band. Notably, we see that there is a positive correlation between band number and information entropy, consistent with the progressive structural refinement observed in high-order bands. This finding suggests that Shannon Entropy provides a useful quantitative design parameters of MCs depending on band orders. In addition, higher-order bands, such as band-6-7 and band-7-8, the band gap value gets dispersive. This finding also suggests that by the combination of entropy estimation and GA exploration, the trends of design rules underlying wide-MBG MCs was extracted via data-driven manner. \par

\section{Conclusion and Outlook}

In this study, we demonstrated the inverse-design topology optimization of 2D MCs combining FD-LLG analysis and genetic algorithm. We efficiently identified several promising Fe-EuO-based bicomponent MCs with nontrivial topological configurations for higher-order bands. Through time-domain micromagnetic simulations, we showed that the topology-optimized structures achieve large CMBG widths of $\Delta f=4.3-5.5$ GHz. Notably, The design optimized for the relative width of CMBG between $n=4$ and $n=5$ exhibits the largest band-gap width ($\Delta f=8.7$ GHz). This design could be particularly promising owing to its structural simplicity and the sufficiently large band gap. Furthermore, the dimensional reduction of MDS further reveals that the design landscape becomes increasingly non-convex in higher-order bands ($n\ge5$). This implies that multiple promising candidate designs exist, and each design has a topology that supports distinctive higher-order magnonic eigenmodes. These results highlight the effectiveness of MC design strategies that expand the design space through high-dimensional structural exploration and exploit higher-order magnonic modes. Beyond conventional optimization based on only a limited number of parameters, such as the bias magnetic field and size parameters (e.g., pitch width and size of scatterer area), optimizing the 2D or 3D topology may open pathway for the design of MCs with greately enhahced flexibility in magnonic band-dispersion engineering. Our proposed framework opens up the possibility of not only discovering unconventional MCs, but also elucidating hidden design principles for the reverse engineered MCs and providing insights for more efficient and robust MC design. Ultimately, this framework may help pave the way for the establishment of unconventional design scheme for magnonic devices for next-generation spintronics applications. \par

\section{Methods}
\subsection{Genetic algorithm}
The process of evolutionary cycle (Selection $\rightarrow$ Crossover $\rightarrow$ Mutation) in genetic algorithm was implemented as follows: \\

	1. $Selection$: This process was performed based on the evaluation results of the objective function. We employed the Roulette Wheel Selection method. In this scheme, the probability of an individual being selected is directly proportional to its fitness value $C_n$ (i.e., the $n$th relative band gap). This implies that individuals with better performance have a higher probability of passing their genes to the next $Crossover$ or $Mutation$ process. The probability $q_i$ of selecting $i$th individual ($1\le i \le P$) in a single generation can be expressed as $q_i=C_n^i/\sum_{j=1}^{P}{C_n^j}$. \\
    
	2. $Crossover$: This operation promotes the recombination of genetic information and facilitates the exploration of the solution space while preserving the structural characteristics of the parents. A "two-point crossover" operator was applied to the parents' gene vectors $\mathbf{x}$ selected in the $Selection$ process, where two crossover points were randomly selected and the segments between them were exchanged between the pair of parent vectors to create children vectors, which is to be treated as parent vectors in the next generation. \\
    
	3. $Mutation$: This process was introduced to maintain genetic diversity and encourage exploration beyond the currently explored region of the search space. Power mutation method was employed, which is a suitable mutation algorithm when the optimization problem has integer constraints \cite{DEEP2009505}. In this method, new children vectors $\mathbf{x}_{\mathrm{c}}$ (parent vectors in next generation) are created from parent vectors inherited from $Selection$ via stochastic process. The $k$th component of children vector is given by:  

\begin{equation}
\mathbf{x}_{\mathrm{c}}=
\begin{cases}
\mathbf{x}(k)-s(\mathbf{x}(k)-\mathbf{l}(k)) & (\mathrm{if} \ d<r) \\
\mathbf{x}(k)+s(\mathbf{u}(k)-\mathbf{x}(k)) & (\mathrm{if} \  d\ge r)
\end{cases}
\label{eq:mutation}
\end{equation}   \\
    
    where $s$ is a random variable extracted from power distribution, $\mathbf{l}$ and $\mathbf{u}$ are arrays of lower and upper bound of the elements, which is $\mathbf{l}=\mathbf{0}$ and $\mathbf{u}=\mathbf{1}$ in this study. $d$ and $r$ are scaled distance of $\mathbf{x}(k)$ from $\mathbf{l}(k)$ defined as $d=\frac{x-l}{u-x}$ and random number extracted from uniform distribution between 0 and 1, respectively. This stochastic variation enables the bit-flip mutation under prescribed probability, allowing algorithm to escape local optima and explore new regions of the search space. \\

\subsection{Entropy calculation}
Shannon entropy was calculated by following equation\\

\begin{equation}
  S=-[x\log_2x+(1-x)\log_2(1-x)],\ x=\frac{V_{\mathrm{EuO}}}{V_{\mathrm{total}}}
  \label{eq:entropy}
\end{equation}\\

Here, $V_{\mathrm{EuO}}$, $V_{\mathrm{total}}$ denotes the volume of the EuO phase and total volume of unit cell, respectively. In the entropy calculation, $2 \times 2$ kernel was prepared, and the entropy within randomly sampled kernels in the unit cell was evaluated $L=10^5$ times, then the averaged value $S_{ave}=L^{-1} \sum_{i=1}^{L} S_i$ was calculated. We employed this quantity to enable the quantification of increasing configurational complexity in MCs along with increasing band numbers.

\section{Acknowledgements}
This work was supported by JSPS Research Fellowships for Young Scientists (DC1) (Nos. 25KJ2117) to R. N. Part of this study was supported by JST ACT-X (Nos. JPMJAX22AL) to T. Y. and JST CREST (Nos. JPMJCR21O1) to M. K. and Y. I.

\bibliographystyle{ieeetr}
\bibliography{references} 

\end{document}